\documentclass[12pt]{iopart}
\newcommand{\ket}[1]{\left| #1 \right\rangle}

\usepackage{graphicx}
\usepackage{ mathrsfs }
\begin{document}

\title[Dynamical decoupling sequence construction as a filter-design problem]{Dynamical decoupling sequence construction as a filter-design problem}

\author{M. J. Biercuk}
\address{Centre for Engineered Quantum Systems, School of Physics, The University of Sydney, NSW 2006 Australia}
\ead{michael.biercuk@sydney.edu.au}

\author{A.C. Doherty}
\address{Centre for Engineered Quantum Systems, School of Physics, The University of Sydney, NSW 2006 Australia}
\ead{andrew.doherty@sydney.edu.au}

\author{H. Uys}
\address{National Laser Centre,\\Council for Scientific and Industrial Research, Pretoria, South Africa}
\ead{huys@csir.co.za}

\begin{abstract}
Over the past decade we have seen an explosion of demonstrations of quantum coherence in atomic, optical, and condensed matter systems.  These developments have placed a new emphasis on the production of robust and optimal quantum control techniques in the presence of environmental noise.  We discuss the use of dynamical decoupling as a form of open-loop quantum control capable of suppressing the effects of dephasing  in quantum coherent systems.  We introduce the concept of dynamical decoupling pulse-sequence construction as a \emph{filter-design} problem, making connections with filter design from control theory and electrical engineering in the analysis of pulse-sequence performance for the preservation of the phase degree of freedom in a quantum superposition.  A detailed mathematical description of how dephasing and the suppression of dephasing can be reduced to a linear control problem is provided, and used as motivation and context for studies of the filtration properties of various dynamical decoupling sequences.  Our work then takes this \emph{practical} perspective in addressing both ``standard'' sequences derived from nuclear magnetic resonance and novel optimized sequences developed in the context of quantum information.  Additionally, we review new techniques for the numerical construction of optimized pulse sequences using the filter-design perspective.   We show how the filter-design perspective permits concise comparisons of the relative capabilities of these sequences and reveals the physics underlying their functionality.  The use of this new analytical framework allows us to derive new insights into the performance of these sequences and reveals important limiting issues, such as the effect of digital clocking on optimized sequence performance.   

\end{abstract}

\maketitle

\section{Introduction}
The needs of the field of quantum science and quantum information have evolved over the past decade.  Just ten years ago the bulk of theoretical and experimental effort in the community was focused on the realization of new quantum coherent systems using condensed-matter devices, optics, and atoms.  The successes of the field~\cite{Ladd2005, Fraval2005, Krojanski2007, Biercuk2009, BiercukPRA2009, Du2009, West2009, Damodarakurup2009, Szwer2010, Alvarez2010, deLange2010, Sagi2010, Bluhm2010, Barthel2010, Cory2010} in this short timeframe have led to the emergence of a growing need for error-resilient quantum control techniques.  Such techniques are enabling for probing studies of the dynamical evolution of quantum systems, the realization of high-fidelity quantum logic operations in quantum information, and the production of useful quantum coherent technologies. 

The ``continuous'' information representation used in quantum technologies means that small deviations in a quantum system's state from the nominally ideal state yields an increased probability of error in a particular basis~\cite{NC2000}.  This makes quantum states particularly sensitive to unwanted Ð and unavoidable Ð environmental couplings, resulting in state randomization through the process of \textit{decoherence}.  Suppressing decoherence has thus become a primary focus of the community as interest shifts away from basic demonstrations of quantum coherence towards attempts at realizing useful quantum coherent technologies~\cite{DVCriteria_1998, 2ndQR}.  

\indent The field of quantum control~\cite{MilburnBook} addresses a broad set of challenges pertaining to the robust and efficient manipulation of quantum systems and processes.  So-called ``open-loop'' control methods, in particular, rely on feedback-free Hamiltonian engineering through the application of time-dependent control sequences, and leverage strengths from longstanding fields such as optimal control theory and nuclear magnetic resonance (NMR)~\cite{Haeberlen1976}.  The latter holds particular importance as modern magnetic resonance imaging (MRI), a biomedical application of NMR, is largely based on pulsed-control techniques designed to overcome the semiclassical analogue of decoherence in a macroscopic ensemble of nuclear spins (for instance in a tissue sample).  The many lessons learned in this field, beginning with Hahn's discovery of the spin echo in 1950~\cite{Hahn50}, can be brought to bear in the challenge of realizing error-resistant hardware for quantum computing applications.
\\
\indent Beginning in the late 1990s significant interest developed in the use of modified concepts from NMR and open-loop control theory for the suppression of decoherence and error in quantum computing applications \cite{Viola1998, Viola1999, Zanardi1999,Vitali1999, Byrd2003,Kofman2004, Khodjasteh2005, Yao2007, Kuopanporttii2008, Gordon2008, Liu2010}.  These protocols, known as Òdynamical decouplingÓ (DD),  are extensions of Hahn's spin echo~\cite{Haeberlen1976, Vandersypen2004}, modified to suit general decoherence processes in qubit systems including the influences of a bath that itself exhibits quantum dynamics.  In these techniques a prescribed sequence of control operations at the physical level enables environmental fluctuations to be coherently averaged out, thus preventing errors from accumulating and preserving quantum information.  The primary degrees of freedom in this approach to quantum error suppression are the number of control operations, $n$, and their relative timing, leading to a wide range of possible DD sequence constructions.    
\\
\indent Altogether, the true power of dynamical error suppression techniques is derived from the facts that: 1) only unitary control operations are employed -- no measurement capabilities nor encoding overheads are required; 2) errors may be suppressed regardless of magnitude, as long as appropriate time-scale separations are obeyed; 3) sequences may be \emph{optimized} for a wide class of noise processes. These facts imply that a sequence of control operations may be applied without an experimentalist having a detailed, real-time quantitative knowledge of the ambient decohering environment, and if applied correctly, an \emph{arbitrary} quantum state of interest will be preserved. 
\\
\indent Using DD it is thus possible to delay the decay of quantum information and suppress decoherence-induced errors to the levels needed for large-scale quantum computation or other applications of quantum systems.  The qubit ``refresh'' techniques to be studied in this work are similar to firmware refreshes of DRAM cells used to compensate for classical information leakage~\cite{Harris2005}.  If quantum information is protected from decoherence in a similar way using dynamical error suppression protocols in support of some ultimate computation or other application, it is appropriate to consider these protocols a form of universally applicable ``Quantum Firmware.''  

In this manuscript we focus on a presentation of the analysis of dynamical decoupling pulse sequences as a problem in filter design, in which we map familiar concepts from electrical engineering such as the 3 dB point, high-pass/band-pass filtering, and low-frequency rolloff (for a high-pass filter) to performance evaluation of various sequences for the problem of pure dephasing.  The objective is to give an accounting of state-of-the-art techniques in dynamical decoupling, but rephrasing the problem in a new framework which has an accessible and practical interpretation, including a quantitative presentation in the context of control theory.  Our work builds on several recent theoretical efforts aiming to explain decoherence and decoherence suppression in terms of spectral functions~\cite{Kofman2004, Martinis2003, Uhrig2007}.  Further, it complements alternative approaches that have been presented recently, using the analogy of optical interference from a diffraction grating to describe dynamical decoupling sequence performance~\cite{Alvarez2011}.   

Using the filter-design perspective we study familiar pulse sequences such as the Carr-Purcell-Meiboom-Gill (CPMG) multipulse spin echo, and Periodic Dynamical Decoupling (PDD)~\cite{Viola1998, Viola1999} and compare against recently discovered sequences such as Uhrig Dynamical Decoupling (UDD)~\cite{Uhrig2007}, Locally Optimized Dynamical Decoupling (LODD)~\cite{Biercuk2009}, Optimized Noise Filtration by Dynamical Decoupling (OFDD)~\cite{Uys2009}, Bandwidth-Adapted Dynamical Decoupling (BADD)~\cite{Viola:BADD}, and others.  In all cases we leverage insights from filter design to concisely explain the performance of the various sequences when applied to the mitigation of decoherence.  Our perspective also permits us to extract new information on the effects of timing constraints on sequence performance, exclusively through study of the modification of the sequence filters.  

Dynamical decoupling can be used to preserve arbitrary quantum states in the presence of general decoherence processes - leading to depolarization.  In this work we focus exclusively on the case of pure dephasing, which while constrained, is of great importance to the community. Further, we do not address concatenated sequences~\cite{Khodjasteh2005} or their performance characteristics in suppressing multi-axis decoherence, and instead focus on the construction of pulse sequences that could form a single concatenation level in a more complex sequence~\cite{UhrigConcatenated, FongPRL2010}. The applicability of the filter-design concepts to more general forms of decoherence remains a subject for future study, but is supported as a general proposition by past theoretical work~\cite{Kofman2004}.

Quantum systems come in many forms, but we will focus on the quantum bit, or qubit, that is most prevalent in quantum information science~\cite{NC2000}.  This abstraction of a quantum mechanical two-level system is useful for information processing, but also serves as a model system for understanding the dynamical evolution of quantum coherent states exposed to noisy environments.  We will frequently employ discussions of technical details derived from quantum information processing as it represents one of the best-known applications for quantum technologies to date.

The remainder of this manuscript is structured as follows.  In Section~\ref{Sec:Error} we review a mathematical formalism for qubit dephasing that is generalized to include the effects of a dynamical decoupling pulse sequence.  This is followed by Sec.~\ref{Sec:FilterDesign} which forms the core theoretical section of the manuscript.  Here, we describe how dephasing can be reduced to a problem in linear control, using insights from control theory, and explain how dynamical decoupling pulse sequences behave as filters in the task of \emph{noise suppression}.  The frequency-domain performance of a given dynamical decoupling sequence is presented in~\ref{Sec:Characteristics}, and we compare the performance of various pulse sequences and discuss relevant operability regimes of interest.  Using these insights, in Sec.~\ref{Sec:FinitePi} we present new analyses of the effects of timing constraints on filter characteristics, including a detailed study of the incorporation of physically realistic (noninstantaneous) control pulses and discretization of time, for instance due to digital clocking.  Sec.~\ref{Sec:Optimization} reviews techniques to create pulse sequences with spectral filtering characteristics optimized for a given noise environment using search algorithms.  The manuscript concludes with a summary and brief future outlook.

\section{\label{sec:Error}Error Accumulation Due to Environmental Noise\label{Sec:Error}}
\indent One of the most pernicious error sources in quantum coherent systems is the phenomenon of decoherence, in which random qubit errors accumulate due to unwanted environmental coupling.  In this section we will describe the effects of decoherence on a qubit system quantitatively, in order to provide physical insight to the dynamical evolution of a quantum state.
\\
\indent The effect of decoherence can be seen by examining the temporal evolution of a qubit initially in a superposition state, e.g.
\begin{equation}
\ket{\Psi}=\cos{\left(\frac{\theta}{2}\right)}\;e^{-i\phi/2} \ket{\uparrow}+\sin{\left(\frac{\theta}{2}\right)}\;e^{i\phi/2}  \ket{\downarrow}\equiv c_{\uparrow}\ket{\uparrow}+c_{\downarrow}\ket{\downarrow},
\end{equation}
where $\ket{\uparrow}$, $\ket{\downarrow}$ are the qubit basis states and $c_{\uparrow,\downarrow}$ are properly normalized, complex probability amplitudes that permit a more compact representation of the state.  In this picture an arbitrary qubit state in this two-dimensional Hilbert space can be represented as a vector on a sphere of unit radius.  This so-called Bloch-sphere representation (lower panel, Fig.~\ref{Fig:schematic}) illustrates how both the relative probability amplitudes and the relative phase between the qubit basis states uniquely define a particular superposition state.   Unwanted environmental coupling has the effect of randomizing the two variables in which state information is encoded, $c_{\uparrow}$ and $c_{\downarrow}$. 
\\
\indent One may consider two general classes of decoherence processes: Longitudinal energy relaxation and transverse dephasing.  In the former process, the probability amplitudes of the qubit state are affected (i.e. $\theta$ or $|c_{i}|$ are randomized), forcing the qubit state to move along a meridian of the Bloch sphere.  The characteristic time over which a two-level system undergoes energy relaxation is known as $T_{1}$~\cite{Vandersypen2004}, borrowing from the NMR literature.  Transverse dephasing involves randomization of the relative phase between the basis states.  In an ensemble average -- for instance, averaging over many different experimental outcomes --  such phase randomization leads to a decay of coherence in a characteristic time $T_{\phi}$.  In total, these two decoherence processes limit the useful lifetime of the system to $T_{2}^{-1}=T_{\phi}^{-1}+(2T_{1})^{-1}$.
\\
\indent The dynamics of a qubit in a general decohering environment may be represented in a semiclassical picture are governed by a Hamiltonian of the form
\begin{equation}
H=\alpha(t)\sigma_{X}+[\frac{\Omega}{2}+\beta(t)]\sigma_{Z},
\end{equation}
where $\sigma_{i}$ represent the Pauli matrices, $\hbar\Omega$ is the unperturbed qubit energy splitting, and $\alpha,\beta$ represent random fields imparted by the environment.  It is generally sufficient to consider only the transverse and one longitudinal process due to the noncommuting nature of the Pauli operators, hence the omission of a term dependent upon $\sigma_{Y}$.  
\\
\indent Application of the time-evolution operator for time $t$ with $\alpha=\beta=0$ leads to a phase accumulation $\Omega t/2$ in the Schrodinger picture.  This phase evolution must always be tracked in a quantum control setting; any arbitrary qubit state accumulates phase due to this term in the Hamiltonian.  It is often convenient, then to write
\begin{equation}
\ket{\Psi(t)}=c_{\uparrow}\ket{\uparrow}e^{i\widetilde{\phi}/2}+c_{\downarrow} e^{-i\widetilde{\phi}/2} \ket{\downarrow}
\end{equation}
where the initial phase of the qubit state at time $t=0$ is absorbed into the $c_{\uparrow,\downarrow}$.  In this way the phase evolution due to Schrodinger's equation is captured explicitly in the variable $\widetilde{\phi}$.
\\
\indent From the perspective of an experimentalist interested in quantum control, qubit coherence is often dominated by transverse dephasing processes.  In a pure dephasing picture $\alpha=0$, $\beta(t)\neq0$, and the state $\ket{\Psi(0}$ evolves to 
\begin{equation}
\ket{\Psi(t)}=e^{-i\int_{0}^{t}\beta(t)dt}\; c_{\uparrow}\ket{\uparrow}+e^{+i\int_{0}^{t}\beta(t)dt}\ \;c_{\downarrow}\ket{\downarrow},
\label{Eq:timeevo}
\end{equation}
where the phase dependent upon $\Omega$ has been dropped by transforming to the rotating frame.  This is represented in the density matrix formalism
\begin{equation}
\hat{\rho}(t)=\left[\begin{array}{cc}|c_{\uparrow}|^{2} & c_{\uparrow}c^{*}_{\downarrow}e^{-2i\int_{0}^{t}\beta(t)dt}\\c^{*}_{\uparrow}c_{\downarrow}e^{2i\int_{0}^{t}\beta(t)dt} & |c_{\downarrow}|^{2}\end{array}\right],
\end{equation}
where the off-diagonal matrix elements give the system coherences, which decay to zero in an ensemble average: decoherence.  For the remainder of this manuscript we shall consider exclusively the case of pure dephasing.
\\
\indent Classical noise in common control or environmental parameters forms a dominant source of dephasing in many quantum control settings; typical dephasing mechanisms include magnetic field fluctuations for atomic systems~\cite{Biercuk2009}, charge fluctuations for charge-qubits in solids~\cite{Fujisawa_PRL_2003}, and the effective Overhauser field arising from randomly fluctuating nuclear spins in semiconductor spin qubits~\cite{Johnson_Nature2005, Reilly_Science_Nuke}.  Experimentalists must also consider, however, the degree to which one may experimentally realize the implicit transformation to the rotating frame.  In a physical sense this transformation corresponds to always keeping track of the phase of the qubit relative to a master oscillator.  We see, therefore, that aside from fluctuations in external parameters, noise in the master oscillator~\cite{DickEffect1, DickEffect2} can also contribute to an effective dephasing between the control hardware and the qubits under test, and that this dephasing may be represented as a contribution of $\beta(t)$~\cite{Biercuk2009}.  
\\
\indent Using these insights we focus on a semiclassical representation of the influence of noise, rather than a full quantum mechanical treatment as may be found in Ref.~\cite{Palma1996, Liu2010}. The ensemble-averaged phase accumulated between the qubit basis states due to random fluctuations in $\beta(t)$ may therefore be equivalently represented as a time-integral over $\beta(t)$ or as a frequency integral over the power spectral density.  
\\
\indent We consider a formulation for measuring the coherence for a dephasing Hamiltonian presented originally by Uhrig~\cite{Uhrig2007, Uhrig2008} and Cywinski~\cite{Cywinski2008}.  Starting with a state $\ket{\Psi_{0}}$ oriented along the $\hat{Y}$-axis in the equatorial plane of the Bloch sphere, the state accumulates a random phase due to environmental noise.  The coherence of this state after time $\tau$ is 
\begin{equation}
\label{Eq:coherence}
W(\tau)=|\overline{\langle\sigma_{Y}\rangle(\tau)}|=e^{-\chi(\tau)}
\end{equation}
\noindent where angled brackets indicate a quantum-mechanical expectation value and the overline indicates an ensemble average.  
\\
\indent Analysis may be conducted in either the time-domain or the frequency domain.  It is often convenient to transform the time-domain noise term $\beta(t)$ to the frequency-domain power spectral density,
\begin{equation}
S_{\beta}(\omega)=\int_{-\infty}^{\infty}e^{-i\omega\tau}\left\langle\beta(t+\tau)\beta(t)\right\rangle d\tau,
\end{equation}
\noindent in which case we may express

\begin{equation}
\chi(\tau)=\frac{2}{\pi}\int\limits_{0}^{\infty}\frac{S_{\beta}(\omega)}{\omega^{2}}F(\omega\tau)d\omega.
\label{Eq:chi}
\end{equation}
\noindent Here $F(\omega\tau)$ describes the spectral dependences of the experiment being performed and is known as the ``Filter Function.''



\subsection{Free-Induction Decay}
\indent The first kind of experiment of interest involves allowing a state to freely precess under the influence of a dephasing Hamiltonian, allowing one to effectively probe the temporal dynamics of the state as in Eq.~\ref{Eq:timeevo}.  After some time $\tau$, the state accumulates a random phase due to the time-integral of $\beta(t)$.  In an ensemble average this produces a decay of coherence, similar to the free-induction decay (FID) experiments of NMR.  Here, we are interested in the total average accumulated phase-error, obtained by integrating over the experimental duration or over the relevant window in frequency space, and thus are only marginally concerned with the spectral content of $S_{\beta}(\omega)$. It has been shown that $F^{(FID)}(\omega\tau)=\sin^{2}\left(\omega\tau/2\right)$~\cite{Martinis2003,Uhrig2008}, and the coherence integral may then be written
\begin{equation}
\chi(\tau)=\frac{2}{\pi}\int\limits_{0}^{\infty}\frac{S_{\beta}(\omega)}{\omega^{2}}\sin^{2}\left(\omega\tau/2\right) d \omega.
\end{equation}

In the limit of low-frequency-dominated fluctuations this expression gives a familiar Gaussian decay envelope to the coherence~\cite{WinelandBible}.  A primary motivation in the following discussion is to extend this coherent lifetime and suppress decoherence-induced errors.

\subsection{Coherence with Dynamical Decoupling}
Dynamical decoupling involves the sporadic application of control pulses in order to average away the effects of coupling between the quantum system of interest and the environment.  We focus on an approach that utilizes the quasi-periodic application of parity-reversing pulses to effectively time-reverse the accumulation of random phase in successive free-precession periods.  More general approaches, including randomized decoupling~\cite{LorenzaRandom}, continuous arbitrary modulation~\cite{Kofman2004, Clausen2010}, or other Pauli rotations~\cite{LaddEcho} may also be considered, but analyzing their construction in the context of filter design will be the subject of future work.

\begin{figure}[tp]
 \begin{centering}
  \includegraphics[width=\columnwidth]{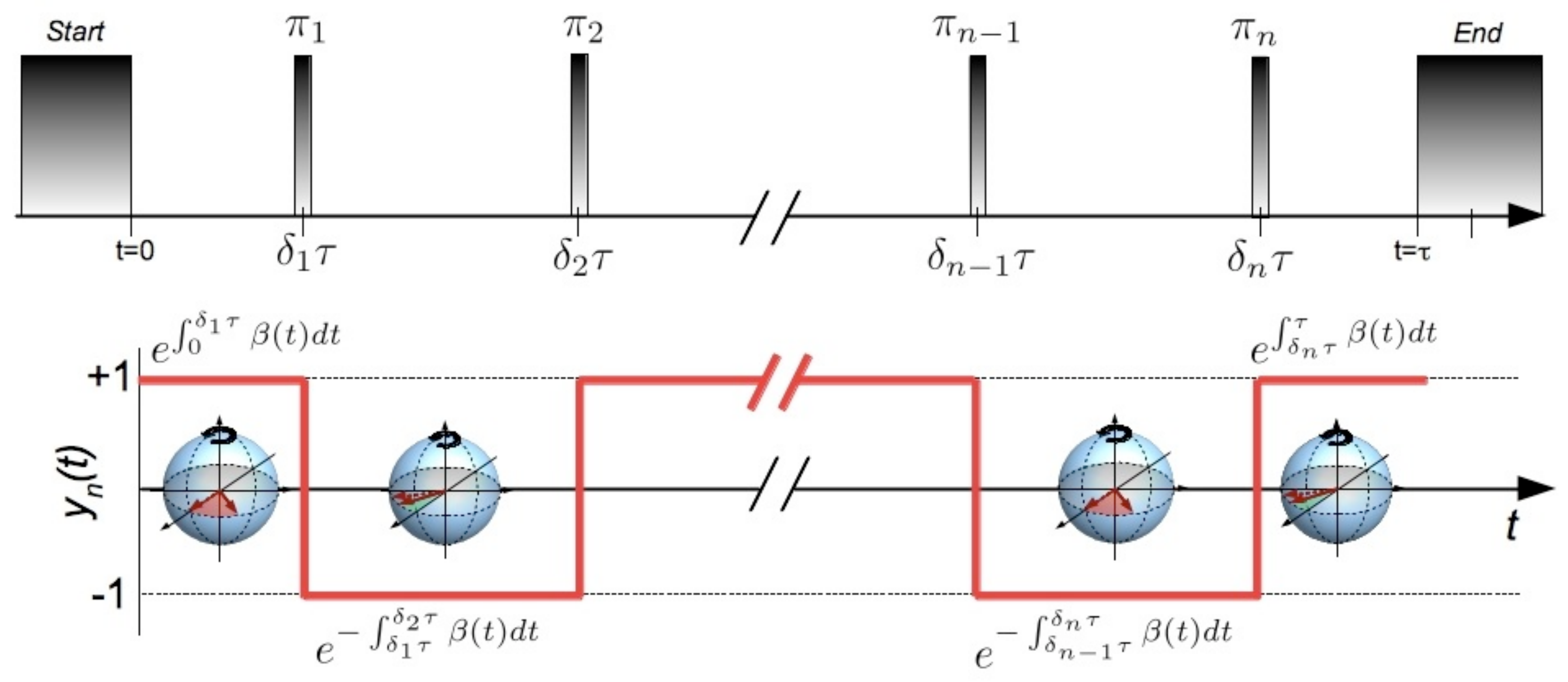}\\
  \caption{\label{Fig:schematic} Schematic representation of a single-axis dynamical decoupling pulse sequence and its relation to the filter function.  (a) Schematic representation of the time-domain application of $\pi$ pulses.  Pulse width is represented as being nonzero for clarity, and are applied about the $\hat{Y}$-axis.  Total sequence duration is $\tau$.  (b) Graphical representation of the toggling frame for phase accumulation corresponding to the integrals in Eq.~\ref{phit}.  Phase accumulation ``winds'' and ``unwinds'' during successive free-precession periods.  Frequency information about the performance of a given dynamical decoupling pulse sequence is captured in the temporal form of $y_{n}(t)$.}
\end{centering}
\end{figure}

Assume arbitrary dephasing-noise as described above and a sequence of $n$ parity-reversing $\pi$-pulses ($\Pi_Y=e^{i\frac{\pi}{2}\hat\sigma_Y}=i\hat\sigma_Y$) during which the instantaneous pulses are applied at fractions $\delta_j$, $j=1,2,... n$, of the total free-evolution time $\tau$. In this case the time evolution of an initial state is given by
\begin{equation}
\ket{\Psi(\tau)}=e^{-i\hat\sigma_Z\int_{\delta_{n}\tau}^{\delta_{n+1}\tau}\beta(t^\prime)dt^\prime}...\Pi_Y e^{-i\hat\sigma_Z\int_{\delta_{1}\tau}^{\delta_{2}\tau}\beta(t^\prime)dt^\prime}\Pi_Y e^{-i\hat\sigma_Z\int_{\delta_{0}\tau}^{\delta_{1}\tau}\beta(t^\prime)dt^\prime}
\ket{\Psi(0)}.
\label{psit}
\end{equation}
In the absence of $\pi$-pulses the total random phase that the qubit accumulates is given by $
\widetilde{\phi}(\tau)=\int_0^\tau\beta(t)dt$ as described in the previous section.  After commuting every other operator $\Pi_Y$ to the left once in Eq.~(\ref{psit}), we see that due to the $\pi$-pulses this random phase becomes instead
\begin{equation}
\widetilde{\phi}(\tau)=(-1)^{n}\int_{\delta_{n}\tau}^{\delta_{n+1}\tau}\beta(t^\prime)dt^\prime... +\int_{\delta_{1}\tau}^{\delta_{2}\tau}\beta(t^\prime)dt^\prime-\int_{\delta_{0}\tau}^{\delta_{1}\tau}\beta(t^\prime)dt^\prime.
\label{phit}
\end{equation}
\noindent In this construction, the state freely evolves, accumulating a random phase, a $\pi$-pulse is applied and the system continues to evolve again accumulating phase, but  \emph{with opposite sign} to that in the first period, and so on.  For an arbitrary time sequence of $\beta(t)$, $t \in [0,\tau]$ the phase evolution of $\ket{\Psi}$ due to the time integral of $\beta(t)$ is modulated by a ``sampling function,'' $y_{n}(t)$, which takes alternating values $\pm1$, with transitions occurring at the temporal-location of the $\pi$-pulses.  Dynamical error suppression therefore involves breaking up the phase-evolution into a series of shorter free-evolution periods in which phase sequentially and repeatedly ``winds up'' and ``unwinds'' (Fig.~\ref{Fig:schematic}), producing an effective averaging of the phase accumulation through the sampling function.

In the case that $\beta(t)$ is static but nonzero over the relevant interval $[0,\tau]$, a single $\pi$-pulse located at time $\tau/2$ can perfectly cancel phase accumulation, constituting the ``spin-echo'' as developed originally by Hahn.  Any variation of $\beta(t)$ between the two free-precession periods reduces the net coherence of the system as the phase accumulation will be slightly different in the two periods.  It is known that by applying a series of $n$ $\pi$-pulses in a particular, evenly spaced sequence, one may form the Carr-Purcell multipulse spin echo (or CPMG given a particular phase relationship between the applied $\pi$ pulses and the qubit state).  Here the phase accumulation is broken up such that in successive free-evolution periods $\beta(t)$ is kept nearly constant, thus maintaining phase-coherence.  Again, variation of $\beta(t)$ between successive free-evolution periods reduces the net coherence of the qubit at the conclusion of the sequence.

Unlike an FID experiment, the performance of a dynamical decoupling pulse sequence is extremely sensitive to the \emph{frequency spectrum} of $\beta(t)$.  An applied dynamical decoupling sequence temporally modulates a qubit's phase accumulation, as presented above; the total free-precession time period is split into smaller time bins separated by $\pi$-pulses.  Slow fluctuations in $\beta(t)$ are corrected via the applied $\pi$-pulses, while more rapid fluctuations between free-precession periods contribute to the net dephasing of the qubit.  The relevant frequency components are set by the total sequence length $\tau$ and the interpulse periods.  As we will see, it will be extremely useful to consider dynamical decoupling pulse sequences as \emph{spectral filters}.

\section{Sequence Construction as Filter Design\label{Sec:FilterDesign}}
The multipulse spin echo (to which we interchangeably refer as CPMG) is a simple dynamical decoupling pulse sequence with a long history of use in NMR.  However it can face significant performance limitations arising from its construction.  As discussed in the last section, the presence of noise that fluctuates rapidly relative to the interpulse spacing diminishes the efficacy of the multipulse spin echo.  In quantum information experiments we aim to maximally preserve coherence rather than simply extend the lifetime of a signal, as in NMR, in the presence of noise characterized by arbitrary power spectral density.   Our present task will be to use the degrees of freedom available to us in order to create new pulse sequences with augmented performance for the specific case of noise which is not well approximated as quasi-static.  We will accomplish this by considering pulse sequence construction as a problem in \emph{filter design}.

\subsection{Filter Basics}
Electrical engineering and digital signal analysis tasks often require an input signal to be mapped to an output signal through a known transformation.  This is frequently enacted in one variable and analyzed in its conjugate.  A prototypical example is the action of changing the spectral characteristics of a time-domain signal, for instance, by removing high-frequency components above a given cutoff.  This approach is generally useful in the removal of extrinsic interference in a signal processing task - for instance, removing mains power frequency modulation of an electrical signal by using a low-pass filter.  In analog electronics, filtering may be achieved using hardware components - for instance networks of passive devices such as resistors and capacitors.  However, given a mathematical representation of a time-signal, a Fourier-domain mathematical filter may also be applied in the case of digital signal processing. For instance, one may perform an FFT on a time-domain signal, remove certain spectral components in software (via the action of the filter), and transform back to the time-domain.

It is instructive to introduce a few basic terms from filter design~\cite{Filters} in order to make plain the corresponding features of classical filtering and the action of dynamical decoupling sequences.  Here, we present a brief description of salient characteristics that is by no means exhaustive.

\begin{itemize}
\item \emph{The Transfer Function}: this describes how a given input signal is transformed to an output signal by the action of the filter.  For a generalized, arbitrary filter we express the application of the filter as the convolution of the input signal with the time domain transfer function of the filter.  This is more commonly expressed through the Laplace transform as $\mathbf{y}_{out}(s)=\mathbf{G}(s)\mathbf{y}_{in}(s)$, where $s$ is a complex variable, and   $\mathbf{G}(s)$ is the transfer function in the fourier domain.  In discrete time this may be represented in matrix form.\\
\item \emph{Frequency Response/ Filter Gain}: used to characterize the aggregate behavior of the Transfer function to a given impulse as expressed in the frequency domain, and describes how particular frequency components of an impulse are modified with either positive (enhancement) or negative (suppression) gain.  For instance, this may include \emph{high-pass}, \emph{low-pass}, \emph{band-pass}, and \emph{notch} behavior.\\
\item \emph{Pass-band and Stop-band}: the regions of frequency space corresponding to transmission and attenuation of relevant frequency components by the filter.\\
\item \emph{Filter Order}: a mathematical parameter used in the construction of the filter which often corresponds to the number of repeated physical units in, e.g. an electrical analog filter.  The filter gain is often conveniently expressed in terms of the filter order.  \\   
\item \emph{Roll-off}: describes the steepness of suppression of a given frequency range below (above) cutoff in a high-pass (low-pass) filter.  Steeper roll-off corresponds to more efficient suppression of noise in the stop-band.  An $\eta$-order filter has $6\eta$ dB/octave = $20\eta$ dB/decade suppression of signals below (above) cutoff.\\
\item \emph{3-dB Point}:   frequency below (above) cutoff at which frequencies are suppressed by 3 dB relative to the pass-band in a high-pass (low-pass) filter.  For practical purposes this is often associated with the cutoff frequency.\\
\item \emph{Ripple}: nonuniformities in the filter gain lead to frequencies being transmitted in the pass-band with varying amplitudes.

\end{itemize}

To understand how the concept of filter design may be applied to dynamical decoupling and quantum control, we must first establish a correspondence between the mathematics of filtering (from the perspective of control theory) and dynamical error suppression.  This requires an explanation of how dephasing due to random noise, characterized by a power spectral density may be reduced to a linear control problem.  The following subsection will use terminology derived from the control theory literature in order to demonstrate the appropriate relationships.  We will then return to the language used more commonly in the physics literature.

\subsection{Dephasing and DD as linear control problems}
\indent The linearity of the dephasing of the qubit may be understood as follows. We saw in the previous section that at the end of a dynamical decoupling sequence that starts at $t-\tau$ and finishes at $t$ we have, before averaging over the noise,
\begin{equation}
\langle\sigma_Y\rangle = {\rm Re}\left\{\exp(- 2 i \tilde{\phi}(t))\right\}=   {\rm Re}\left\{\exp\left(- 2 i \int_{(t-\tau)}^0 h(t-t')\beta(t') dt'\right)\right\}.
\end{equation}
\noindent Here $h$ is the impulse response function implicitly defined in equation (11),  which defines the filter that is applied to the noise $\beta$ to produce the phase $\tilde{\phi}$, and is equivalent to the sampling function $y_{n}(t)$ for a chosen pulse sequence.  Writing things in this way emphasizes that there is a linear dynamical map from the noise $\beta$ to the phase of the coherent qubit oscillation, and suggests the use of techniques from filtering and control. It is important to note though that it is not possible to generate a general impulse response $h$ from the sorts of dynamical decoupling pulses given in equation (11) so the class of possible linear evolutions is constrained. \\
\indent When we average over the noise we find our figure of merit
\begin{equation}
W= e^{-\chi(\tau)} = e^{- \langle \tilde{\phi}^2(\tau)\rangle}
\end{equation}
and it is clear that we can optimize our pulse sequence of length $\tau$ if we minimize $\langle \tilde{\phi}^2(\tau)\rangle$. We regard this as our {\it objective function} to be minimized. Intuitively $h$ should be designed so that $\tilde{\phi}$ does not respond sensitively to $\beta$. In control theory language, $\tilde{\phi}$ is not very {\it controllable} by the noise input $\beta$.

\indent In control theory there is a standard description of this sort of problem as follows~\cite{RobustControl}: consider a physical system with states described by a vector ${\bf x}$ undergoing linear dynamics and driven by some control or noise signals given by ${\bf u}$ that produces some output ${\bf y}$ that is a linear function of ${\bf x}$
\begin{eqnarray}
\dot{\mathbf{x}}=A\mathbf{x}+B\mathbf{u}(t)\\
\mathbf{y}=C\mathbf{x},
\end{eqnarray}
\noindent which may be solved to give
\begin{eqnarray}
\mathbf{x}(t)=e^{At}\int_{0}^{t}e^{-At'}B\mathbf{u}(t')dt'\\
\mathbf{y}(t)=Ce^{At}\int_{0}^{t}e^{-At'}B\mathbf{u}(t')dt'.
\end{eqnarray}
\noindent Applying the Laplace transform we find
$\mathbf{y}(s)=\mathbf{G}(s)\mathbf{u}(s)$, where
$\mathbf{G}(s)=C\left(sI-A\right)^{-1}B$ is the transfer function for
the control operation of interest, $I$ is the identity, and $s$ is a
complex number.

In our system we simply have one input $\mathbf{u}(t)=\beta(t)$ and one output $\mathbf{y}(t)=\mathbf{x}(t)=\tilde{\phi}(t)$. We describe our dynamical decoupling experiment as follows. We consider our state to be initialized at time
$t=-\infty$ and evolve unperturbed until time $t=-\tau$.  At this time
the Hamiltonian term $\beta(t)=B\mathbf{u}(t)$ is turned on and the
state allowed to evolve until time $t=0$, at which point the state
evolution is said to be complete.  In a free evolution process we set
$\mathbf{G}(s)=I$, and the state evolves under the application of the
dephasing Hamiltonian.
\\
\indent When considering the prevention of dephasing we are dealing
with a linear control problem in which we are attempting to implement
\emph{noise suppression}.  Our system experiences a disturbance
defined by $\beta(t)=B\mathbf{u}(t)$, and our task, in applying a
dynamical decoupling pulse sequence, is to filter out the effect of
the noise, i.e. to \emph{reduce} the system's controllability by the
environment.  As a result of our objective function we do not care about values of the output for $t<0$, only the phase at the end of the pulse sequence matters.  As a result we are interested in the map from the input noise signal $\mathbf{u}(t)$ for $t<0$ to the state at $t=0$,
$\mathbf{x}(0)$.  This mapping is linear, and in control theory is known as the {\it controllability operator}. We write
\begin{equation}
\mathbf{x}(0)=\mathscr{C}\left[\mathbf{u}(t)\right],
\end{equation}
\noindent where we can think of the controllability operator $\mathscr{C}\left[\mathbf{u}(t)\right]$ as an
asymmetric matrix that represents the transform of $\mathbf{u}$ under
the action of transfer function $\mathbf{G}(s)$.  The expectation value of interest
for our control problem is 
$\left\langle\mathbf{x}^{T}(0)\mathbf{x}(0)\right\rangle=\left\langle
\mathbf{u}^{T}\mathscr{C}^{T}\mathscr{C}\mathbf{u}\right\rangle$,
which we seek to minimize. (The angle brackets here indicate averaging over realisations of the noise.) This formula is analagous to the construction of
time evolution under a dephasing environment and an open-loop control
protocol to be represented by the quantum mechanical operator
$\hat{\Theta}$, where the expectation value is represented at time
$t=0$ as $\left\langle\Psi(0)\right|\hat{\Theta}\left|\Psi(0)\right\rangle={\rm Tr}\left(\Theta\ket{\Psi(0)}\left\langle\Psi(0)\right|\right)$. This suggests some further simplifcation of the formula for the objective function.
We may
write,

\begin{eqnarray}
\left\langle\mathbf{x}^{T}\mathbf{x}\right\rangle&=\left\langle
\mathbf{u}^{T}\mathscr{C}^{T}\mathscr{C}\mathbf{u}\right\rangle\\
&=\left\langle {\rm Tr}\left(\mathscr{C}^{T}\mathscr{C}\mathbf{u}\mathbf{u}^{T}
\right)\right\rangle\\
&={\rm Tr}\left(\mathscr{C}^{T}\mathscr{C}\left\langle
\mathbf{u}\mathbf{u}^{T}\right\rangle \right)\\
&={\rm Tr}\left(\mathscr{C}\mathbf{V}\mathscr{C}^{T} \right).
\end{eqnarray}
\noindent Here the quantity $\mathbf{V}$ denotes the covariance matrix
of $\mathbf{u}$, playing a role directly comparable to
$S_{\beta}(\omega)$.  In the case of Gaussian white noise this reduces
to the identity, however this quantity is nontrivial in the case of
non-Markovian processes which exhibit colored noise (e.g. $1/\omega$).
The quantity $\mathscr{C}\mathbf{V}\mathscr{C}^{T}$ is in general a matrix, known to control theorists as the {\it weighted controllability Grammian},  and provides all information about how noise with a colored spectral representation, under a control procedure (here dynamical decoupling) maps $\mathbf{x}(-\infty)\rightarrow\mathbf{x}(0)$.  (Weighted because of the effect of $\mathbf{V}$ to account for the spectrum of the noise). For our class of problems we would wish to minimize the eigenvalues of the controllability Grammian by choosing the transfer function $\mathbf{G}$ such that the system does not respond strongly to the noise. In our case the weighted controllability Grammian is simply the number $\chi$.

\subsection{DD in the Frequency Domain}
\indent The control-theoretic task in a dynamical decoupling experiment using continuous (discrete) time can be reduced to construction of a $F(\omega\tau)$ ($\mathbf{G}$, or $\mathscr{C}$) that maximally suppresses noise characterized by $S_{\beta}(\omega)$ ($\mathbf{V}$).  

\indent The time-domain modulation due to a decoupling sequence enacts the desired \emph{noise suppression} through the convolution of $\beta(t)$ and  $y_{n}(\tau)$.  Hence we look to the modulation's Fourier transform to provide the relevant spectral information.  In particular we may write the filter function~\cite{Uhrig2007,Uhrig2008},
\begin{equation}
F(\omega\tau)=|\tilde{y}_n(\omega\tau)|^{2}=|1+(-1)^{n+1}e^{i\omega\tau}+2\sum\limits_{j=1}^n(-1)^je^{i\delta_j\omega\tau}|^{2} 
\label{Eq:FFfull}
\end{equation}
\noindent This quantity provides all information about how an arbitrary pulse sequence will suppress phase accumulation as a function of frequency, and exists in analogy to the \emph{filter gain}.  The filter function enters our expression for coherence, $W(\tau)=e^{-\chi(\tau)}$, as described in Eq.~\ref{Eq:chi}. Since the filter function appears as a multiplicative factor of $S_{\beta}(\omega)$, small values of $F(\omega \tau) $ in the dominant spectral regime will lead to small values of $\chi(\tau)$, and hence coherence $W(\tau)\approx1$.  We see from this presentation that the control-theoretic $\mathscr{C}^{2}\propto\frac{1}{\omega^{2}}F(\omega\tau)$ from the physics literature.  

\indent Examination of Eq.~\ref{Eq:FFfull} indicates that changing $n$ or modifying the fractional pulse locations $\delta_{j}$, in an $n$-pulse sequence can alter its frequency response.  The fact that $\delta_{j}$ is an easily controlled parameter has produced a recent focus in the community on the construction of new DD sequences tailored to suppress dephasing noise with arbitrary spectral characteristics by modifying the form of the filter gain.  

A sequence design methodology based on modification of the filter function has provided a new avenue for the construction of experimentally relevant sequences and has allowed a simple physical insight into the action of dynamical decoupling broadly. An experimental operator may construct a dynamical decoupling sequence to provide a $F(\omega \tau)$ that is most useful for the noise present in the experiment -- one that minimizes the objective function defined above.  

This approach is taken routinely when an experimentalist attempts to improve qubit coherence by adding pulses to a CPMG sequence;  doing so reduces the minimum pulse separation for a given total experimental duration and hence allows correction for higher-frequency (more rapidly changing) components of $S_{\beta}(\omega)$.  Recent theoretical investigations~\cite{Khodjasteh2005,Uhrig2007,Uhrig2008,Lee2008, Cywinski2008} have demonstrated that substantial improvements in achievable memory fidelities in the presence of rapidly fluctuating system baths (and sharp high-frequency cutoffs) are possible by resorting to more sophisticated sequence design, as will be described below.  

Equivalent to the control theory perspective above, the frequency-domain filter function shows how an applied dynamical decoupling pulse sequence behaves explicitly as a \emph{spectral filter}, with gain characteristics captured in the mathematical form of $F(\omega\tau)$.  For low frequencies the phase-factors in the filter function construction are small compared to $\pi/2$ and the terms add destructively to produce the stop-band.  In this regime the filter function remains small and noise contributions to $\chi(t)$ are minimized.  For the highest frequencies the phase-factors can differ by $\pi$, permitting constructive interference and yielding rapid fluctuations about the mean value $4n+2$.  Frequency components of $S_{\beta}(\omega)$ in this regime may contribute substantially to $\chi(t)$, yielding dephasing and primarily impacting $T_{2}$.  In this way we may think of a dynamical decoupling sequence as behaving like a \emph{high-pass filter}.  The filter function in the pass-band exhibits significant ripple and can take values greater than unity, producing effective enhancement of dephasing due to noise at certain frequencies.

We define the analog of a classical electrical filter's 3-dB point as the crossover between the two regimes described above; it is the value of $\omega\tau$ (or $\omega$ for a given $\tau$) giving a value of the filter function $F_{n}\sim1$, denoted $\omega_{F1}$.  This corresponds approximately to the ``dynamical decoupling limit,''~\cite{Viola:BADD} 
\begin{equation}
\frac{\omega}{2\pi}<\frac{1}{\tau_{min}},
\end{equation}
\noindent where $\tau_{min}$ is the minimum interpulse period.  This says that dynamical decoupling will successfully permit rephasing so long as the highest-frequency component of $S_{\beta}(\omega)$ is slow relative to the smallest interpulse period.  Returning again to the filter function formalism, if $S_{\beta}(\omega)$ has significant components $\omega>\omega_{F1}$, the system will be largely dephased as the applied sequence cannot decouple the qubit from the rapidly fluctuating environment, i.e. the filter gain passes these frequencies approximately unimpeded.

In many experimental studies of coherence, $\tau$ is varied and the system's coherence is measured.  Because $\tau$ enters the expression for the filter function and sets the frequency-response characteristics of a given dynamical decoupling sequence (along with $\delta_{j}$ and $n$), variation in $\tau$ will change the frequency response of the sequence.   Thus the accumulation of error will not necessarily result in a simple exponential or even Gaussian decay envelope in $\tau$ when dynamical decoupling pulses are applied.  Instead, the form of the decay can be highly irregular for a particular sequence even showing ``revivals'' where the measured error occasionally decreases with increasing $\tau$ for certain ranges, depending upon the characteristics of the noise~\cite{Biercuk2009}.

\section{\label{Sec:Characteristics}DD Filter Characteristics}
\subsection{Sequences of Interest}
Early DD schemes have largely relied on the simple periodic repetition of (approximately) instantaneous pulses.  The best known among these are the CPMG (after Carr Purcell, Meiboom, and Gill, and referenced above) and Periodic Dynamical Decoupling (PDD) sequences.  

The CPMG sequence was originally developed in the context of NMR systems where inhomogeneous broadening required an ability to refocus the ensemble Bloch vector as it spread out during free precession \cite{Haeberlen1976}.  However, the same sequence is quite effective at suppressing phase randomization due to homogeneous effects (e.g. $\beta(t)$) when noise processes are dominated by low-frequency components (e.g. $S_{\beta}(\omega)\propto1/\omega$).  An $n$-pulse CPMG sequence is constructed with fractional pulse locations $\delta_{j}=(j-1/2)/n$ (Fig.~\ref{Fig:schematic}a) wherein the first and last free-precession periods are half the duration of the interpulse periods, producing effective ``refocusing'' of the Bloch vector at the conclusion of the sequence.   

Periodic Dynamical Decoupling involves the repetitive application of uniformly spaced $\pi$ pulses with fractional locations $\delta_{j}=j/(n+1)$ for $n$ pulses~\cite{Viola1998}.  The sequence does not provide efficient refocusing of the Bloch vector, but does perform averaging of environment-induced phase accumulation due to its sampling function~\cite{Cywinski2008}.  However, this sequence has the benefit of providing suppression of general decoherence, if properly constructed.

Uhrig first showed that manipulation of the relative pulse locations, $\delta_{j}$, for fixed $n$ and $\tau$ leads to modification of $F(\omega\tau)$, providing the ability to tailor the filter function.  In particular, he analytically derived an $n$-pulse sequence in which the first $n$ derivatives of $\tilde{y}_{n}(\omega\tau)$ vanish for $\omega\tau=0$.  Expressed alternatively, given an expansion of $\beta(t)$ in powers of $t$, $\beta(t)=\beta_{0}+\beta_{1}t+\beta_{2}t^{2}+\beta_{3}t^{3}+...$, the UDD sequence is constructed to suppress the first $n$ orders of the expansion, given $n$ pulses, while the CPMG sequence cancels \emph{exactly} only the zeroth and first-order terms for all $n>2$~\cite{BiercukQIC2009, Szwer2010}.  The resulting sequence, UDD, has $\pi$ pulse locations determined analytically as $\delta_{j}= \sin^{2}[\pi j/(2n+2)]$.
\\
\indent The construction presented above allowed Uhrig to tailor the filter function such that it provided strong suppression of phase accumulation when noise environments possessed significant high-frequency contributions --- a dramatic advance over CPMG.  Uhrig specifically showed \cite{Uhrig2008} that in noise spectra including high-frequency components and a sharp high-frequency cutoff, $\omega_{D}$, such as the Ohmic spectrum ($S_{\beta}(\omega)\propto\omega\Theta(\omega_{D}-\omega)$) that may be consistent with a spin-boson model~\cite{Leggett1987, Uhrig2007, Kofman2004}, the UDD sequence would yield significant gains in performance relative to CPMG~\cite{Uhrig2007, Uhrig2008}.  By contrast, in the presence of noise with a soft high-frequency cutoff, the error-suppression benefits arising from the form of the filter function for UDD were reduced.  

\subsection{Regimes of interest}
It is important to distinguish, at this point, between two salient regimes of operability in quantum systems.  We note that a distinction between these regimes has not been formalized previously.
\begin{enumerate}
\item ``High-fidelity'' regime, occurring at times $t<<T_{2}$, where $T_{2}$ is the $1/e$ coherence time of the system.  In this regime the accumulated error due to decoherence is small, but must be compared against, e.g. predicted fault-tolerance error thresholds of $p_{th}\approx0.01\%$ derived from quantum error correction.  In quantum computing applications the maximum allowable error must not surpass $p_{th}$, which is a much more stringent requirement than the typical experimental metric that $\tau<T_{2}$. 
\item ``Coherence time'' regime, in which error probabilities of a few tens of percent are permissible, and $T_{2}$ is the only relevant metric.  This may be appropriate for, e.g. new quantum enabled sensing or imaging applications where large error probabilities provide contrast between different targets.  
\end{enumerate}

\subsubsection{The filter function in the high-fidelity regime}
\begin{figure}[b]
 \begin{centering}
  \includegraphics[width=\columnwidth]{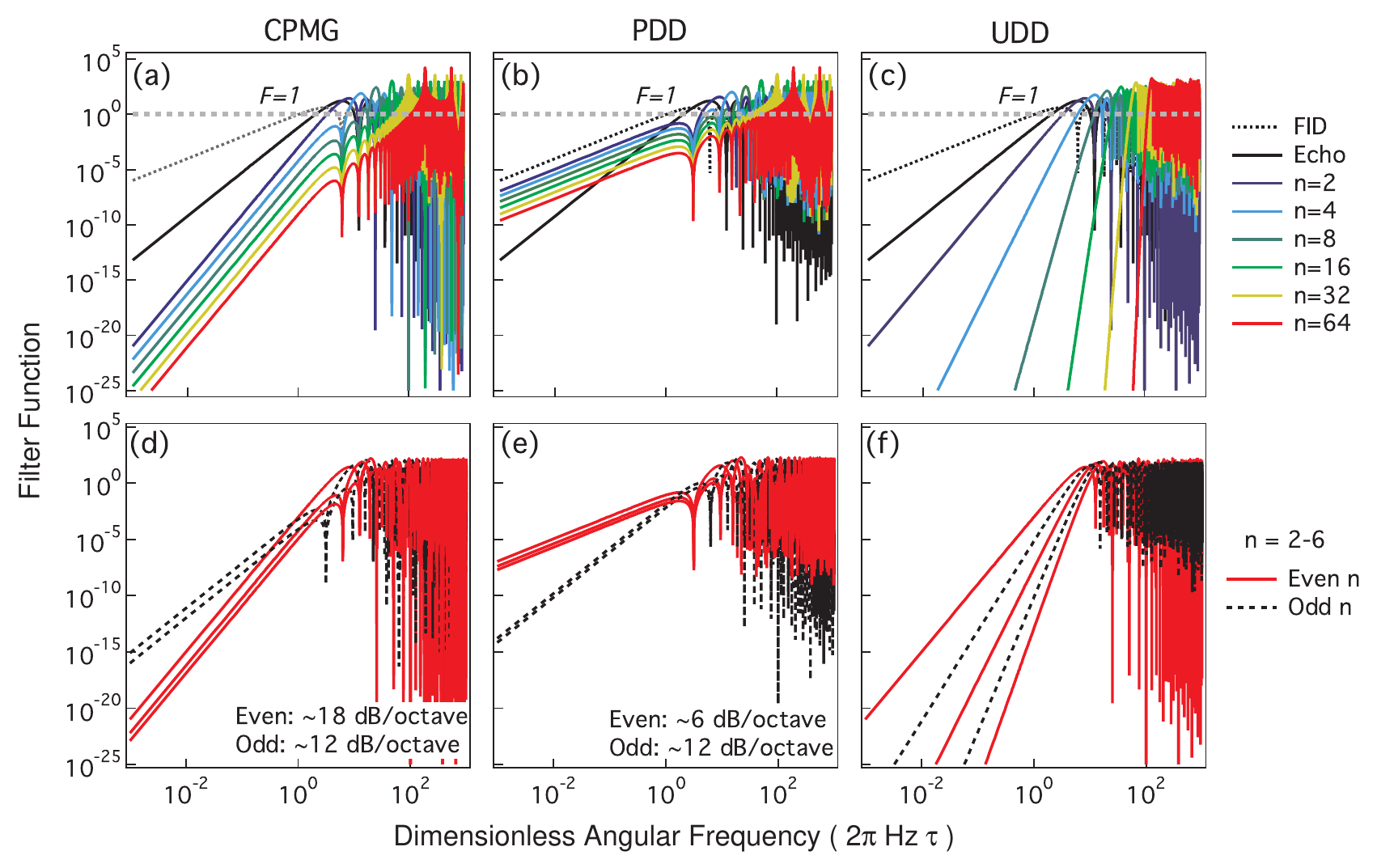}\\
  \caption{\label{fig:FF1} Graphic representation of the filter function for various pulse sequences as a function of dimensionless angular frequency ($\omega\times\tau$).  (a-c) Numerical filter function for various $n$ and pulse sequences.  Dashed grey line indicates the numerical value $F=1$ above which noise is not suppressed by the filter function.  (d-f) Filter functions for the same sequences and $n=2-6$ demonstrating even-odd asymmetry in noise suppression of the various filter functions, as manifested in the slope of the low-frequency rolloff or the floor value in the case of UDD.}
\end{centering}
\end{figure}

The filter function interpretation of DD appropriate for the high-fidelity regime is shown graphically in Fig.~\ref{fig:FF1} where the filter function of multiple $n$-pulse DD sequences is presented on a log-log plot for various values of $n$.  In these plots we have normalized to a dimensionless angular frequency ($\omega\times\tau$), as this is the only physical quantity that enters into Eq.~\ref{Eq:FFfull}.  In all cases, the filter function is small at low frequencies and grows with frequency.  Near $\omega_{F1}$ it begins to oscillate rapidly.   

Examination of panels (a-c) of this figure show that the low-frequency rolloff in the stop-band varies dramatically between sequences, providing significant performance differences between these sequences for a given $S_{\beta}(\omega)$.  Using the framework of filter-design we can analyze the filter functions for various sequences using new quantitative metrics.  The low-frequency rolloff of PDD for these values of $n$ is $\sim$6 dB/octave while for the same $n$ CPMG has a rolloff of $\sim$18 dB/octave.  This can be compared to the difference between the performance of single-pole and multi-pole filters in electrical engineering.  As such, low-frequency noise is filtered much more efficiently using CPMG, where the filter function can be more than 100 dB lower than that for PDD given the same $n$.  

The value of $\omega_{F1}$ increases approximately linearly with $n$, indicating an increase in the spectral range efficiently filtered by the applied sequence.  In neither case, however, does increasing pulse number change the low-frequency rolloff, i.e. the effective \emph{order} of the filter.

For both CPMG and PDD we observe an even-odd asymmetry (Fig.~\ref{fig:FF1}d, e), but the pulse-number parity providing the best low-frequency rolloff differs between the sequences~\cite{LaddThesis, Haeberlen1976}.  In the case of PDD, even pulse numbers provide the worst suppression of low-frequency noise, comparable in form ($\sim$6 dB/octave) to that of FID.  This is commensurate with the observation that given an even pulse number, in PDD we have an odd number of equal-length free-precession periods.  As such, phase accumulation due to quasi-static noise is not fully compensated following the final free-precession period, making the system more susceptible to low-frequency noise.  While this asymmetry has been recognized in the physics community, its origin is elucidated via the filter-design framework.

The performance of the UDD sequence is quite distinct from the filtering characteristics of CPMG and PDD (Fig.~\ref{fig:FF1}f).  Most importantly, the UDD sequence provides a filter function whose rolloff increases in steepness with $n$.  This is a manifestation of the analytic condition Uhrig originally imposed, in which the first $n$ derivatives of $F^{(UDD)}(\omega\tau)\equiv0$.   As such, the addition of pulses in the UDD sequence is analogous to increasing the filter order in a standard filter design problem.  By contrast, the low-frequency rolloff for CPMG and PDD is approximately constant with $n$, limiting their flexibility and performance.  Additionally, even-odd parity differences in the UDD filter functions are not visible in the slope of the low-frequency rolloff immediately below $\omega_{F1}$.

\begin{figure}[b]
 \begin{centering}
  \includegraphics[width=8cm]{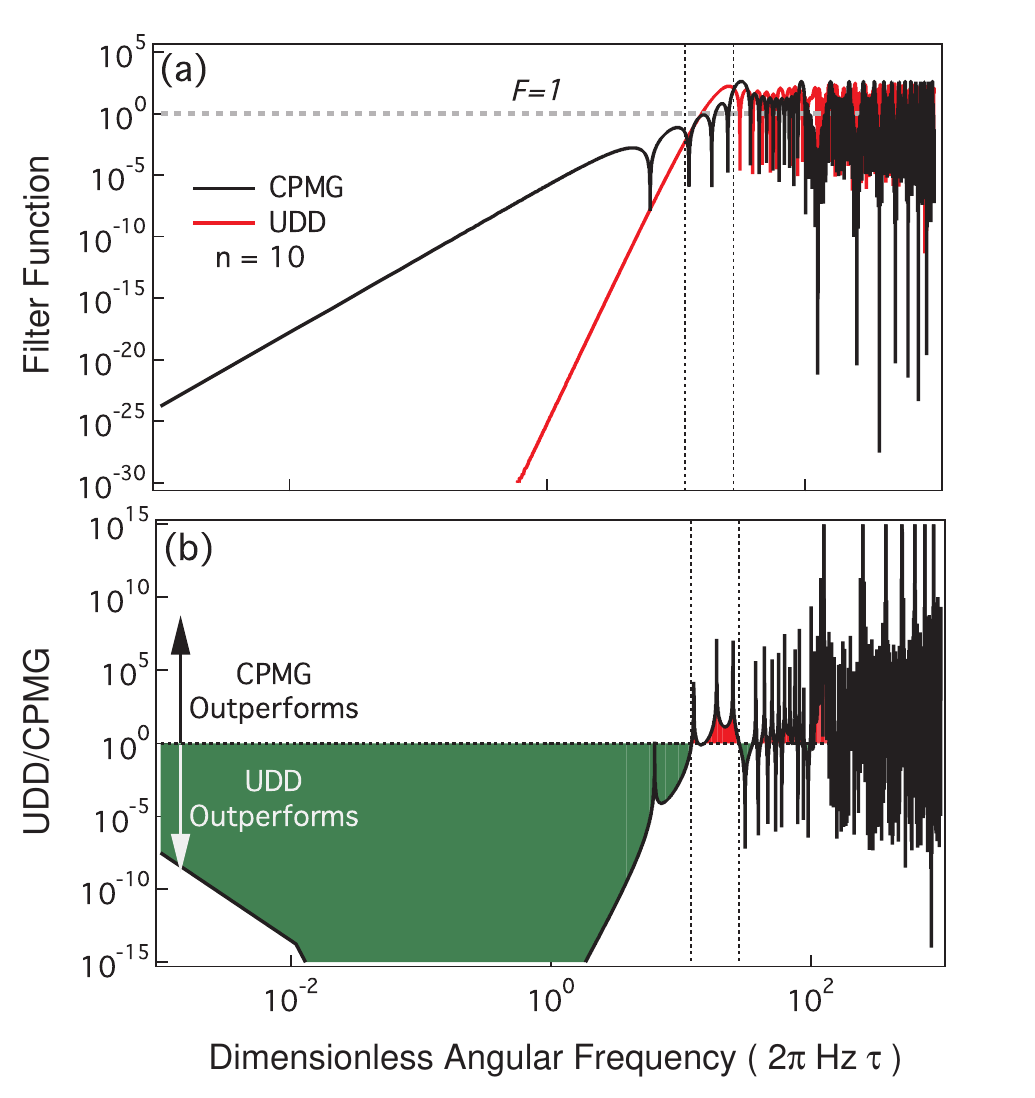}\\
  \caption{\label{fig:FF2} Comparison of the CPMG and UDD filter functions.  (a)The filter functions calculated numerically for CPMG and UDD, with the value unity indicated by the dashed grey line. (b) The numerical value of the UDD filter function divided by that for CPMG.  A value of one indicates equal filter functions, values greater than one correspond to CPMG outperforming UDD (red shading), and vice versa for values less than one (green shading).  Dashed lines in both panels indicate the frequency range near $F=1$ where CPMG outperforms UDD, generally giving $T^{(UDD)}_{2}<T^{(CPMG)}_{2}$ in the presence of non-negligible noise spectral weight in this region.  The benefits of UDD at the lowest frequencies are artificially limited as both filter functions saturate below $\sim10^{-35}$ due to numerical limitations.}
\end{centering}
\end{figure}

Differences in the ultimate error-suppressing capabilities of CPMG and UDD are clearly illustrated by a comparison of the filter functions for  fixed $n$.  In Fig.~\ref{fig:FF2}a we plot the filter functions for $n=10$ for CPMG and UDD, and the ratio of the filter functions in Fig.~\ref{fig:FF2}b.  We clearly observe that for the majority of the spectral range of interest (below $\omega_{F1}$), the filter function for UDD can be over 15 orders of magnitude smaller than that for CPMG, providing superior suppression of decoherence due to the associated components of $S_{\beta}(\omega)$.  A broad spectral range of noise is ``maximally'' filtered by UDD before the onset of a sharp reduction of filter performance as $\omega\rightarrow\omega_{F1}$, an insight previously presented in various works.  

Based on these calculations, we see that for most instances, complex optimized sequence construction can provide significant benefits in noise suppression in the high-fidelity regime.  UDD can outperform CPMG by many tens to hundreds of dB over experimentally relevant ranges of $S_{\beta}(\omega)$, and given a sufficiently sharp high-frequency noise cutoff (corresponding to the sharp turn-on in the UDD filter function near $\omega_{F1}$), UDD generally gives near-optimal error suppressing performance in this regime.

\subsubsection{The filter function in the coherence time regime}
The tradeoff between spectral filtering range and filtering efficiency is also manifested in the observation that in most instances, for a given $n$, $\omega^{(UDD)}_{F1}<\omega^{(CPMG)}_{F1}$.  We observe this in Fig.~\ref{fig:FF2}b as a region in which CPMG outperforms UDD.  Noise with significant power spectral density in the frequency range denoted by the dashed lines contributes to dephasing largely unimpeded as this region is close to $\omega_{F1}$ (Fig.~\ref{fig:FF2}a), or its effect may even be \emph{amplified} by the presence of positive gain in the filter functions.  As this noise contributes large dephasing-induced error probabilities, it predominantly impacts the measured value of $T_{2}$ in the system.

\begin{figure}[b]
 \begin{centering}
  \includegraphics[width=9cm]{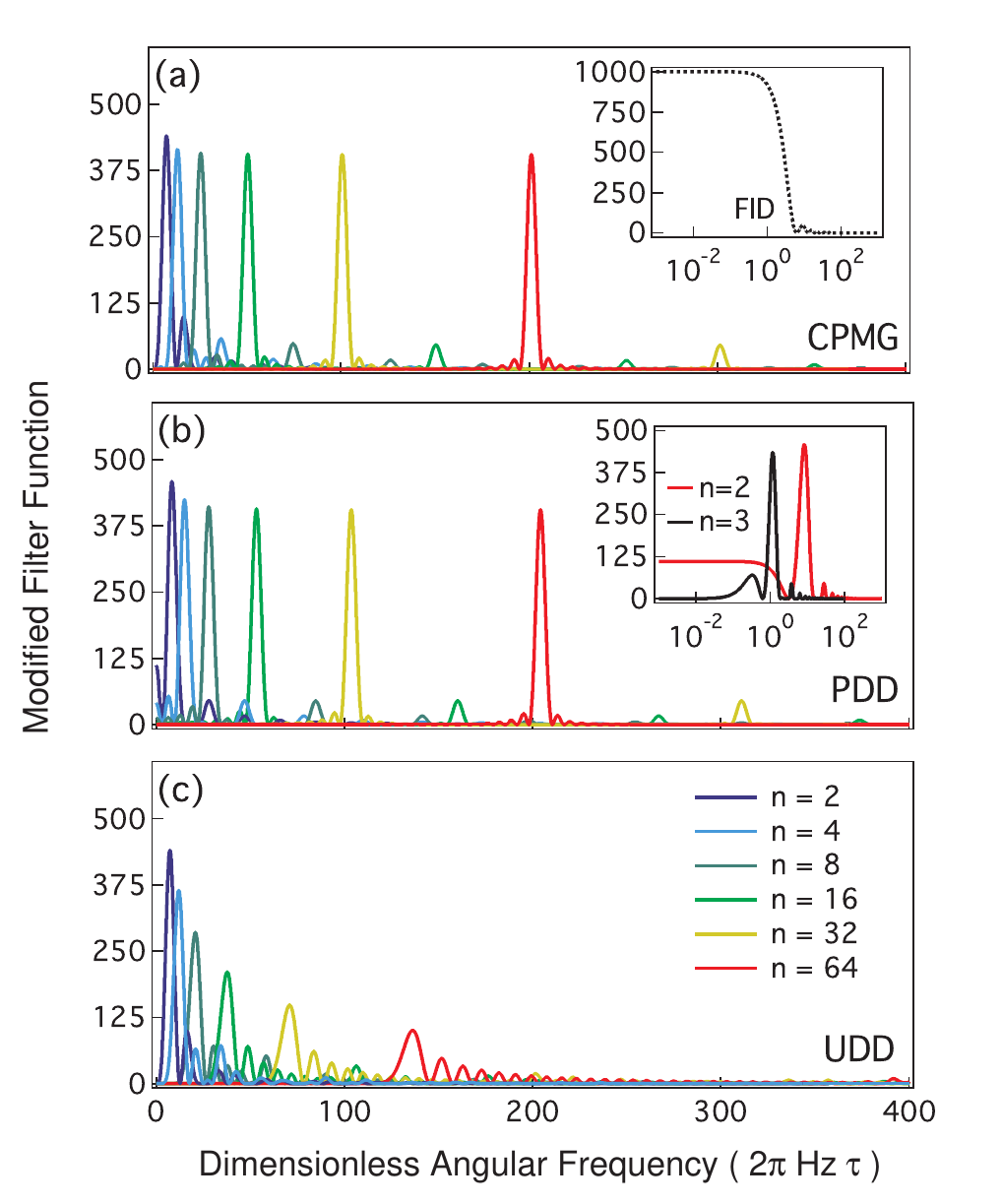}\\
  \caption{\label{fig:FF3} (a-c )Modified filter function, $F^{(T_{2})}(\omega\tau)$, for the pulse sequences and values of $n$ studied in Fig.~\ref{fig:FF1} on a linear scale.  Dominant spectral contributions to the measured $T_{2}$ appear as peaks in the modified filter functions.  Inset a) Modified filter function for FID showing large weight for low-frequency noise on a semilog scale, with arbitrary units.  Inset b) Demonstration of even-odd parity through the modified filter function of PDD on a semilog scale.  }
\end{centering}
\end{figure}

Contributions to the system $T_{2}$ under dynamical decoupling sequence application may be clarified by wrapping the $1/\omega^{2}$ term in Eq.~\ref{Eq:chi} into the expression for the filter function~\cite{Oliver_PC}, yielding a modified filter function $F^{(T_{2})}(\omega\tau)=F(\omega\tau)/\omega^2$. This construction emphasizes the parts of $F(\omega\tau)$ near $\omega_{F1}$ that provide the largest dephasing-error contributions and dominate $T_{2}$.  

The modified filter functions for CPMG, PDD, and UDD are plotted on a linear scale in Fig.~\ref{fig:FF3}, where pass-band ripple produces a peak in the modified filter function.  This spectral peak serves as an approximate \textit{band-pass} filter for noise dominating $T_{2}$, with a minimum of $\sim$20-26 dB of rejection outside the passband.  Again, the pulse-sequences \emph{always} serve as high-pass filters, but by focusing exclusively on the noise components that dominate the measured $T_{2}$ we find an approximate narrowband sampling of $S_{\beta}(\omega)$.  Note that for noise sufficiently dominated by low-frequency contributions, spectral components near $\omega_{F1}$ will not necessarily dominate the measured $T_{2}$ for a given sequence.

For CPMG and PDD we observe narrow spectral bands shifting towards higher frequencies with increasing $n$.  This is derived from the fact that the use of equal interpulse free-precession times corresponds to selection of a fixed frequency component in the Fourier transform of the sampling function; in this case approximately a squarewave function.   Low-frequency components are filtered out by the action of the pulses.  This may be contrasted against the modified filter function for FID (Inset, Fig.~\ref{fig:FF3}a) that demonstrates significant contributions from noise at the lowest frequencies, corresponding to extreme sensitivity to, e.g. shot-to-shot quasistatic fluctuations in an ensemble average measurement of error in a FID experiment.  The low-frequency behavior of even-pulse-number-parity filter functions for PDD is demonstrated in the inset to panel (b).  Like FID, low-frequency components are sampled broadband for even pulse numbers, equivalent to the discussion above about uncompensated quasistatic noise.  Increasing the pulse number decreases the low-frequency constant value of the modified filter function.  This is again expected as increasing $n$ reduces the uncompensated phase accumulation time during the final free-precession period, thus reducing sensitivity to quasi-DC noise.  These familiar behaviors are explained succinctly in the filter-design formalism.

UDD again presents a departure from the observations for CPMG and PDD.  In this case, increasing pulse number corresponds to a shift in spectral peak towards higher frequencies, but also a broadening of the spectral range contributing appreciably to $T_{2}$.  This arises through both a widening of the primary spectral peak with $n$, and also the emergence of higher-frequency oscillations above the primary peak.   These features correspond to the slight increase in the standard filter function around $\omega_{F1}$ with $n$ as depicted in Fig.~\ref{fig:FF1}c, and the fact that the sampling function of the UDD pulse sequence contains a variety of spectral components (corresponding to each of the different interpulse precession times).  Broadband noise can therefore lead to a shortening of $T^{(UDD)}_{2}$ relative to $T^{(CPMG)}_{2}$, even if error rates at short times are lower for UDD than for CPMG.

Interestingly, this form of the filter function has recently been employed in a pair of exciting experimental demonstrations.  First, an experiment by Bylander and colleagues showed how the use of CPMG could permit tunable sampling of the noise power spectral density in a form of noise spectroscopy~\cite{Oliver_PC}.  Second, Kotler et al demonstrated a ``quantum lock-in amplifier'' in which dynamical decoupling suppresses broadband noise but permits detection of a desired modulation signal chosen in the passband of a PDD sequence~\cite{Roee2011}.  Our studies illustrate how, in the modified-filter function interpretation, periodic pulse sequences will provide the most narrow pass-band spectral response, while other sequences optimized for operation in the high-fidelity regime will sample noise broadband in contributing to $T_{2}$.  Thus a new ``optimization'' procedure will be needed for bandpass applications of DD sequences.

\section{Effects of Nonzero $\tau_{\pi}$ and other Timing Constraints\label{Sec:FinitePi}}
\indent Almost all theoretical studies of the effects of dynamical decoupling have been performed in the so-called ``Bang-Bang'' limit~\cite{Viola1998}, in which control-pulse operations are assumed to be delta-functions, with pulse duration $\tau_{\pi}=0$.  This assumption is convenient theoretically, but unphysical as any real system will have nonzero $\tau_{\pi}$.  The problem of accounting for realistic pulses \cite{Viola2003, Zhang2008} has attracted considerable attention of late, and has yielded important insights.
\\
\indent Returning to the filter function formalism we may make the lowest order approximation accounting for nonzero $\tau_{\pi}$ by modifying the sampling function $y_{n}(t)$ to incorporate a delay of length $\tau_{\pi}$  and value zero between values of $\pm1$ corresponding to the free-precession periods~\cite{Biercuk2009, BiercukPRA2009}.  This approximation assumes that environmentally induced phase accumulation is negligible during the application of a $\pi$ pulse, consistent with many experimental observations in which Rabi decay times (the decay of driven oscillations) are much longer than FID times.  When moving to the frequency domain, incorporating this delay results in a filter function for an arbitrary $n$-pulse sequence
\begin{equation}
F(\omega\tau)=|1+(-1)^{n+1}e^{i\omega\tau'}+2\sum\limits_{j=1}^n(-1)^je^{i\delta_j\omega\tau'}\cos{\left(\omega\tau_{\pi}/2\right)}|^{2},
\end{equation}
where $\delta_{j}\tau'$ is the time of the center of the $j^{\rm th}$ $\pi_{X}$ pulse, and $\tau'=\tau+n\tau_{\pi}$ is the sum of the total free-precession time and $\pi$-pulse times.  The underlying assumption of zero dephasing during applied pulses breaks down as $\tau_{\pi}\rightarrow\tau$.  In this limit, the system's evolution is dominated by periods of driven rotation rather than free-precession and evolution of $\beta(t)$ during extended pulses prevents the cancellation of phase accumulation in successive free-precession periods.

\begin{figure}[tbp]
 \begin{centering}
  \includegraphics[width=11cm]{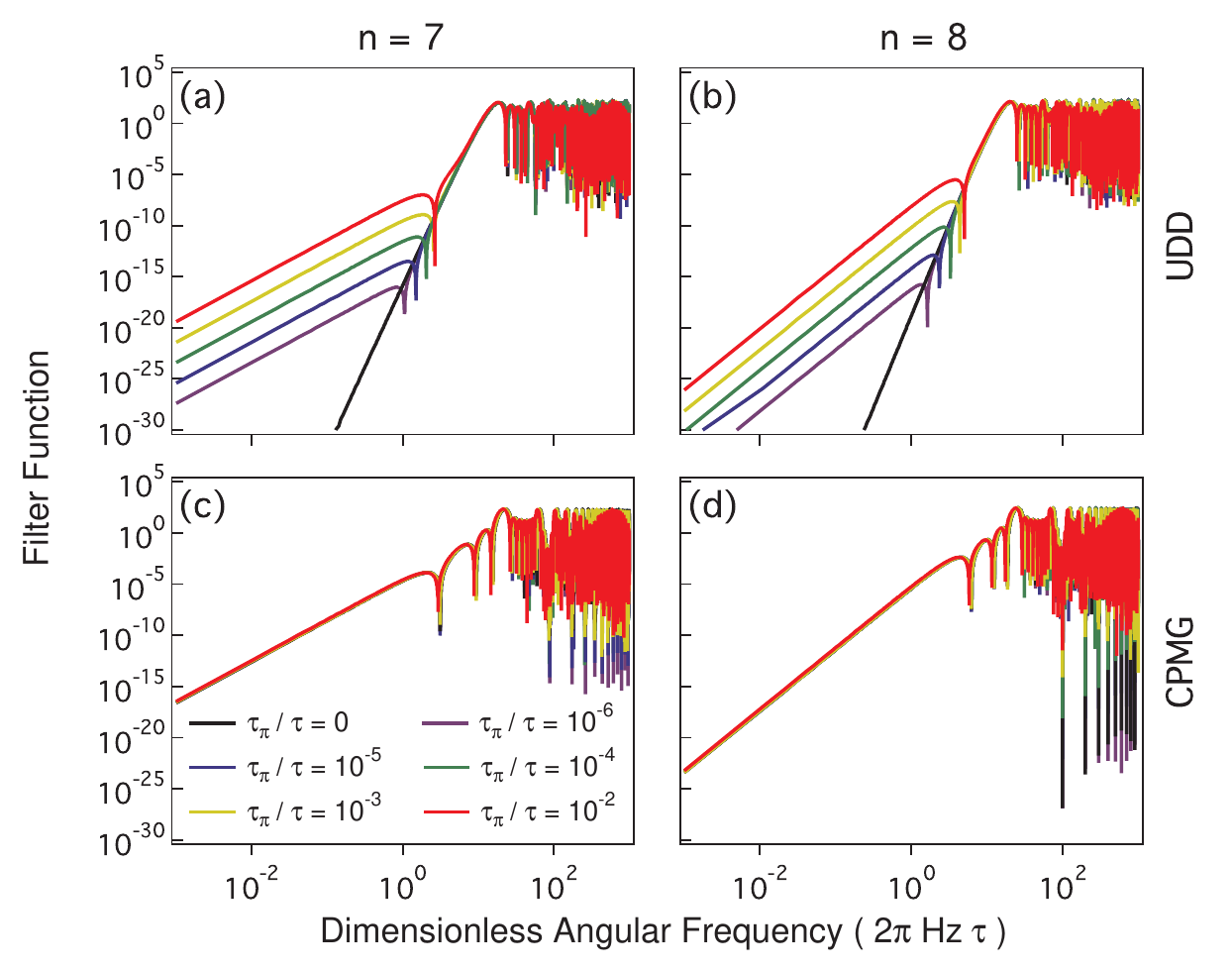}\\
  \caption{\label{Fig:FF4} Modification of the filter function by inclusion of nonzero $\tau_{\pi}$ for UDD and CPMG with $n=7$ and $n=8$.  In these panels different traces correspond to different values of the ratio of $\tau_{\pi}$ to the total free-precession time, $\tau$.}
\end{centering}
\end{figure}

\indent The inclusion of such a delay in the sampling function, and the corresponding modification of the filter function is represented graphically in Fig.~\ref{Fig:FF4} for UDD and CPMG.  Including a nonzero $\tau_{\pi}$ in this manner most significantly effects the UDD sequence in which the roll-off of the filter function is reduced by accounting for $\tau_{\pi}$.  As $\tau_{\pi}$ increases, the roll-off of the UDD filter function converges to approximately the same slope as that for CPMG, eliminating the benefits of the more complex sequence construction, and identified for the first time here.  For the regime studied in this figure, CPMG is minimally affected by the modification to the filter function. Here, $\tau_{\pi}$ is approximately six times smaller than the shortest interpulse free-precession period, $\tau^{(CPMG)}_{min}$, but as $\tau_{\pi}$ becomes comparable to this quantity, $\omega_{F1}$ is shifted to lower frequencies (not shown).  PDD behaves similarly to CPMG.  While general modification of the filter function via inclusion of nonzero $\tau_{\pi}$ has been verified by experiment~\cite{Biercuk2009, BiercukPRA2009}, the new analyses of dynamical decoupling sequences as filters present significant new insight into the relative importance of $\tau_{\pi}$ for various sequences.

\begin{figure}[tbp]
 \begin{centering}
  \includegraphics[width=11cm]{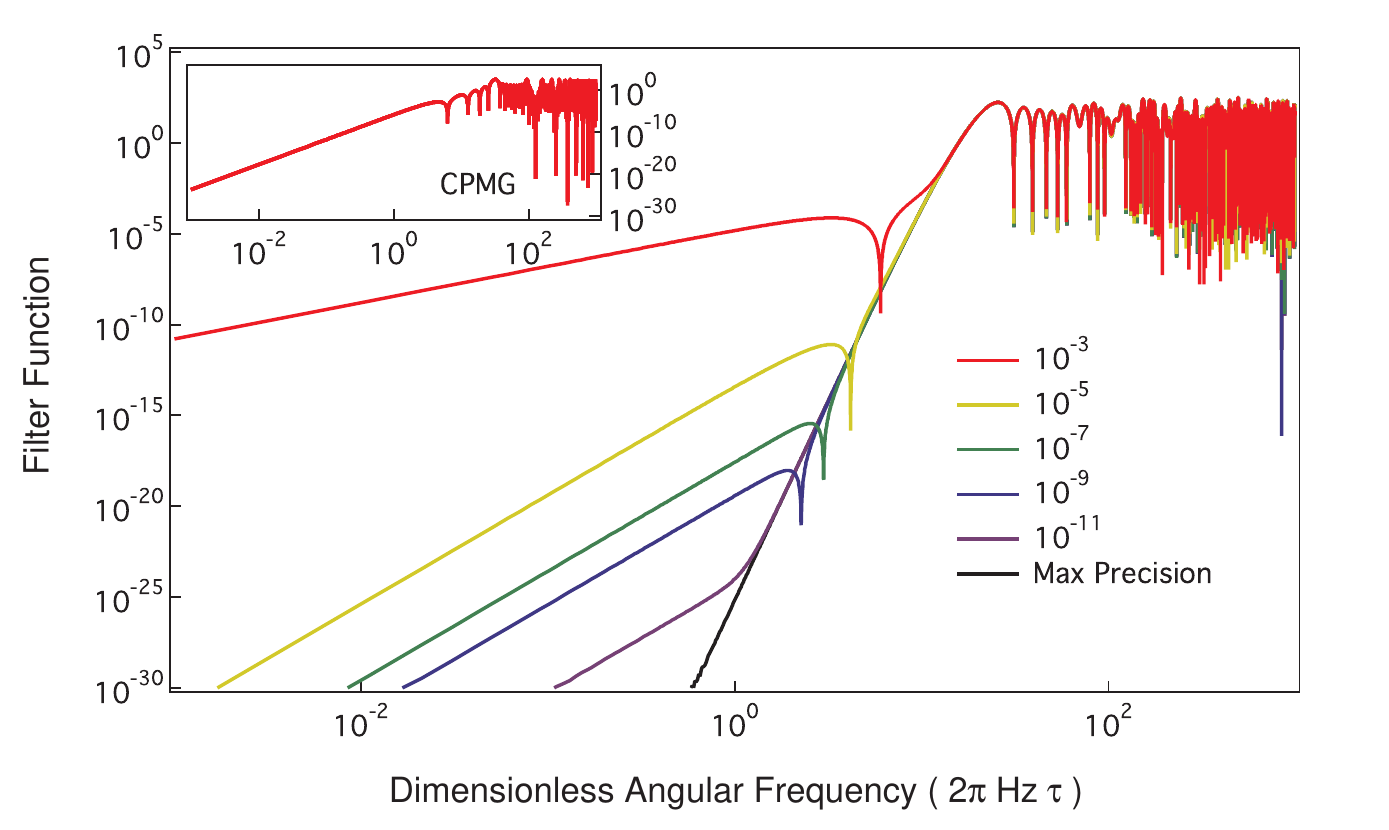}\\
  \caption{\label{Fig:FF5} The effect of imprecision in pulse location. The vector $\delta_{j}$ is numerically rounded to the precision indicated in the legend.  Maximum precision corresponds to double-precision integer representation.  Here $n=10$ and the main panel shows the filter function for UDD.  Inset) Filter function for CPMG; effects of rounding are not visible up to $10^{-2}$}
\end{centering}
\end{figure}

A new analysis indicates that similar effects can be identified by reducing the precision with which pulse locations are specified (Fig.~\ref{Fig:FF5}).  We may construct a filter function using instantaneous pulses but impose rounding on the fractional pulse locations used to produce the filter, effectively discretizing time and corresponding, e.g. to the effects of digital clocking.  In particular, rounding of pulse timing accuracy to just one part in $10^{7}$ can reduce the error suppression of the UDD filter function at a dimensionless angular frequency of unity by $\sim80$ dB, with increasing losses in suppression at lower frequencies.  This could correspond to the effects of a 10 GHz clock (100ps clock period) on an experiment with duration 1 ms.  Again, we observe that CPMG is nearly unaffected by the imposition of these constraints to the level of $\sim1\%$.  The discretization of time discussed here may actually provide benefits in the analysis of filter design, borrowing from the control theory literature.

Ultimately, this understanding of the filter function suggests that realistic hardware constraints may impose a nontrivial limit in the efficacy of optimized dynamical decoupling sequences such as UDD.  Previous experiments have shown that such timing constraints could be ignored on the level of error rates $\sim10^{-3}$, but may become appreciable as error rates are pushed near fault-tolerance limits~\cite{BiercukPRA2009}.   

\section{Filter Optimization through Search Algorithms\label{Sec:Optimization}}
 Each dynamical decoupling sequence discussed hereto had a  specific filtering design objective; in the case of spin echo and CPMG that objective was to cancel slowly varying noise terms, while an $n$-pulse UDD sequence is constructed to cancel the first $n$ orders in a Taylor expansion in time of any classical noise field~\cite{Szwer2010}.   These sequences have in common that they define a set of \textit{relative} inter-pulse delays which is  the same for arbitrary total sequence lengths.  One might ask whether, by abandoning this philosophy, the filter design problem can be recast to find sequences that are either universally optimal or tailored to outperform all other sequences.  We will see that it is possible to construct new  \emph{suites} of sequences, specifically tailored to a given noise environment in order to provide maximum error suppression.

\subsection{Locally Optimized Dynamical Decoupling (LODD)}
The first optimization approach involves designing a different noise filter for each value of the total sequence time, $\tau$, as this parameter sets the overall frequency characteristics of a given filter function (Eq.~\ref{Eq:FFfull}).   By definition the best noise filter is that one which minimizes the coherence integral in Eq.\ref{Eq:chi}.  The corresponding analytical condition is that the derivative of the coherence integral with respect to each pulse location, $\delta_j$, must vanish, which leads to a set of coupled equations to solve.  Since a solution to that set of equations is not possible for arbitrary $S_\beta(\omega)$ it is necessary to resort to numerical means.  This philosophy was employed in \cite{Biercuk2009, BiercukPRA2009} where the coherence integral was directly minimized for a given $S_{\beta}(\omega)$ (as opposed to solving the set of coupled equations) via a multidimensional search algorithm.  The resulting sequences were referred to as LODD sequences (Locally Optimized Dynamical Decoupling) in reference to the unique sequence obtained locally for each $\tau$.  

\begin{figure}
\centering
\includegraphics[scale=0.75]{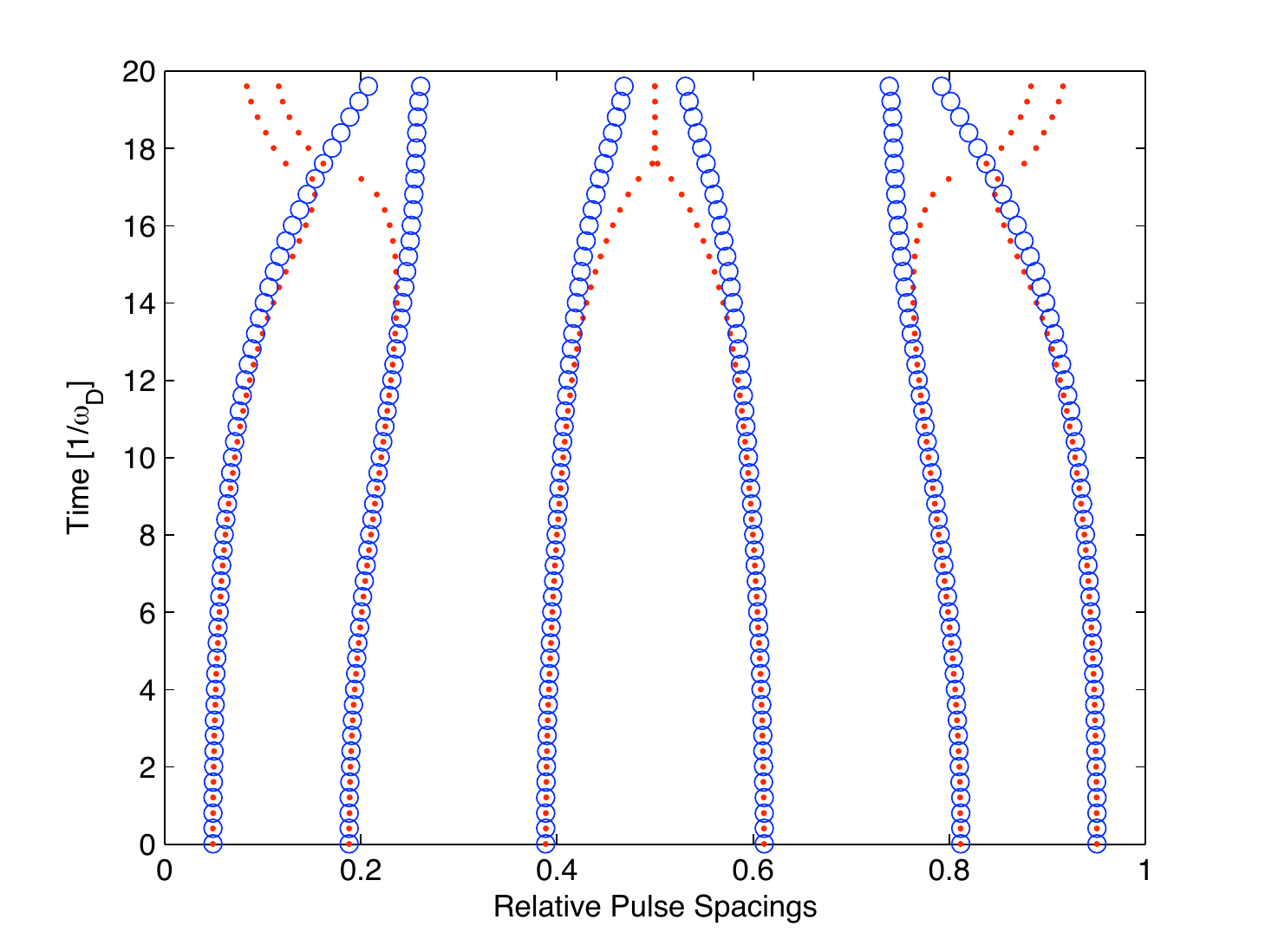}
\caption{Relative pulse spacings of LODD (open circles) and OFDD (diamonds).  The relative pulse spacing multiplied by the total sequence length, $\tau$, provides the pulse locations in absolute terms.}
\label{LODDOFDDtimings}
\end{figure}

A set of such sequences, with $n=6$, is shown in Fig.~(\ref{LODDOFDDtimings}) where the open circles indicate the relative timings of the LODD sequence for each time indicated on the vertical axis.  In this case the sequences were optimized for an ohmic spectrum $S(\omega)=\omega$ with a sharp frequency cutoff at $\omega_D$.  As with UDD, the LODD approach yields significant improvements when the noise spectrum is high-frequency dominated and has a sharp high-frequency cutoff~\cite{Biercuk2009}.  Moreover LODD was shown to provide significant gains over UDD across a range of $\tau$ under those circumstances.

An advantage of the LODD approach is that it can be {\it experimentally} implemented using closed-loop feedback without any prior detailed knowledge of the power spectrum on the part of the operator; minimization of measured error is the only task to be carried out.   This is accomplished by initiating a search for the optimized sequence with a random guess and measuring the resulting coherence.   A new guess is then generated using a suitable search algorithm and the coherence remeasured.  The process is repeated until the algorithm converges within the resolution afforded by the measurement fidelity.  


\subsection{Optimized Noise Filtration Dynamical Decoupling (OFDD)}
The noise power spectrum enters directly into the numerical calculation of LODD sequences through the coherence integral, $\chi(\tau)$.  It is therefore reasonable to expect that the LODD sequences found would be very different for different noise spectra.  One study has shown that locally optimized sequences can be found that are suitable for use under many circumstances \cite{Uys2009}. Consider a new optimization objective of minimizing the area under the filter function, Eq.~\ref{Eq:FFfull}, up to some appropriate frequency cutoff, $\omega_D$.  In an experimental system that cutoff will be set by a characteristic frequency of the system such as a spectral bandwidth of the noise, but need not be known {\it a priori}. Since the coherence integral is determined by the overlap between the filter function and noise spectrum, minimizing the area under the filter function will minimize the overlap without concern for detailed information about $S_{\beta}(\omega)$.   

Again, a suite of pulse sequences is obtained, but this same set of sequences many be applied to a broad range of noise spectra.  The pulse sequence at each total sequence length, $\tau$, is unique and predetermined numerically. This approach was referred to as optimized noise filtration dynamical decoupling (OFDD).  In an experiment a feedback search algorithm need only perform a one-dimensional search through the predetermined set of pulse sequences to find that one yielding the best coherence for a specific $\tau$, constituting a significantly reduced experimental-feedback burden.  This then sets the optimum sequence for every other $\tau$.  Figure \ref{LODDOFDDtimings} compares the pulse spacings for LODD optimized for the ohmic spectrum to that of the OFDD sequences in the same units.  In the high-fidelity regime these sequences differ minimally.  
\begin{figure}
\centering
\includegraphics[scale=0.75]{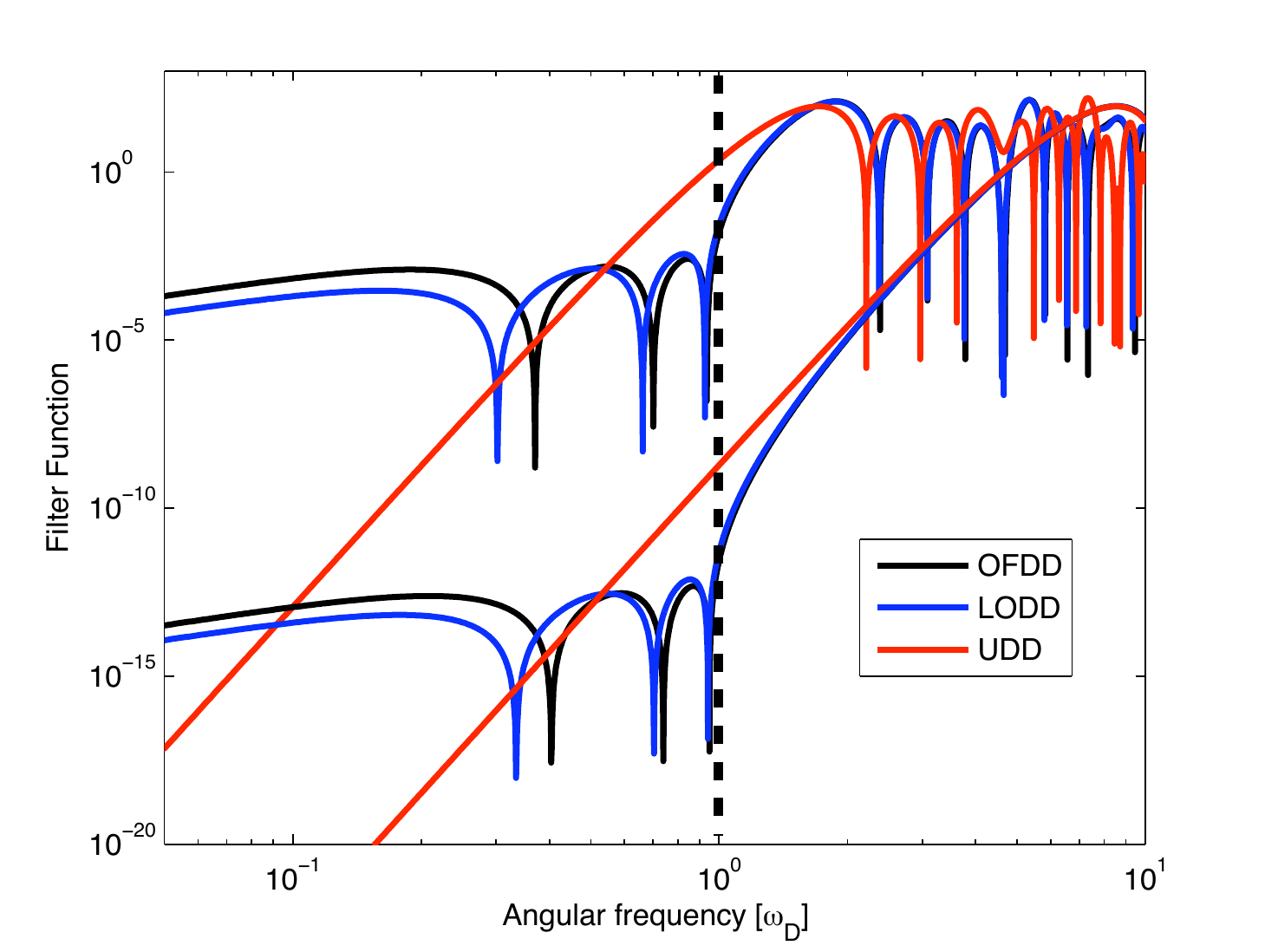}
\caption{Comparison of filter functions for UDD (red), LODD (blue) and OFDD (black) sequences at $\tau=5$ and
$\tau=10$.  The dashed line corresponds to the imposed cutoff frequency, $\omega_{D}$.}
\label{LODDOFFDfilters}
\end{figure}

Reference \cite{Uys2009} indicated that for certain noise environments the LODD and OFDD approaches can significantly improve on sequences in which the relative pulse spacings are the same for all sequence times, while doing no worse than other sequences in most environments. To gain further insight to why this is so it is instructive to compare the filter functions of locally optimized sequences directly. Figure \ref{LODDOFFDfilters} plots the filter functions for UDD (red), LODD (blue) and OFDD (black) sequences with $n=6$ and at two different times $\tau=5$ and $\tau=10$ (units of $1/\omega_D$). LODD was optimized for the same ohmic spectrum as in Fig. \ref{LODDOFDDtimings}.  Note that LODD and OFDD are qualitatively very similar.  In both cases the filter functions are larger in magnitude than that due to UDD in the low-frequency regime, providing somewhat weaker noise suppression over this spectral range. Nonetheless these sequences perform better than UDD because they increase the slope of the low-frequency roll-off and hence suppress noise more strongly close to the high-frequency cutoff, i.e. that region where the filter function is usually largest, and where for an Ohmic spectrum the noise is strongest.

\subsection{Bandwidth Adapted Dynamical Decoupling (BADD)}
\indent In experimental settings control technologies impose constraints on DD often neglected in theoretical studies, such as imperfect detection, pulse errors, finite energy constraints, and finite switching time of the control modulation.  Incorporating such control constraints into an optimization routine can produce novel sequences which improve dephasing supression.  For instance, the LODD sequence construction described above implicitly accounts for $\tau_{\pi}$ by simply finding the suite of pulse sequences minimizing $\chi(\tau')$.  A more explicit accounting for timing constraint is addressed by Khodjasteh et al., in the formulation of Bandwidth-Adapted Dynamical Decoupling (BADD)~\cite{Viola:BADD}.

There, Khodjasteh {\it et al.} \cite{Viola:BADD} re-evaluated the numerical DD optimization problem in light of finite timing resources; they imposed the constraint that the shortest inter-pulse delay, $\tau_{min}$,  in any DD sequence must obey:
\begin{equation}
\tau_{min} \geq \tau_{switch},
\end{equation}
where $\tau_{switch}$ is a minimum switching time allowed by the control technology.  They then performed numerical searches for optimum pulse sequences in which both $\tau_{min}$ and the total sequence time $\tau$ are constrained, but $n$ is allowed to vary.  The resulting sequences were referred to as \textit{bandwidth-adapted} DD (BADD) sequences.   With these timing constraints UDD sequences are much more severely limited in the number of allowed pulses for a given $\tau$ than the BADD approach.

The authors compare the decoupling error for UDD and BADD sequences in a spin-boson dephasing model with a supra-Ohmic power spectrum $S_{\beta}(\omega)=\alpha\omega^3\exp{(-\omega/\omega_c)}$, where $\alpha\approx 1.14\times 10^{-26}$  s$^2$ and $\omega_c\approx 3$ rad/ps.  There a timing constraint is of $\tau_{min}\approx 0.1$ ps is imposed by the need to avoid unwanted excitation of higher energy levels.  Then for a total sequence time of $\tau\approx10^2$ ps BADD and LODD sequences allow up to $n=100$ while UDD can only accomodate $n=20$ commensurate with the constraint.  The authors observe that the minimum achievable error is only very weakly dependent on the total sequence time for the BADD sequences.  This is an indication that when operating in a regime of `fast-control'  DD performance should be measured in terms of $\tau_{switch}$ rather than $\tau$, to allow fair comparisons. UDD on the other hand performs significantly poorer due to the $n=20$ limit.


\section{Conclusions\label{Sec:Conclusions}}
In this manuscript we have discussed the construction of dynamical decoupling pulse sequences for the efficient suppression of decoherence as a problem in filter design, analogous to tasks in digital signal processing or analog electronics.  This approach provides a \emph{practical} perspective on the functionality of various dynamical decoupling pulse sequences, and helps to elucidate the relative performance advantages and limitations of certain optimized constructions.   This work has provided a comprehensive review of recent results in the field, but has also yielded new results derived explicitly from the use of a filter-design analytical framework.

We have analyzed the performance of the leading pulse sequence formulations including PDD, CPMG, and UDD, and reviewed new optimization techniques such as LODD, OFDD, and BADD entirely in the context of filter design.  A quantitative analysis derived from classical control theory forms the theoretical foundation for this treatment, and is based upon the reduction of dephasing-suppression to a problem in linear control theory.  Our analyses have yielded new insights into how these sequences provide robust performance in particular noise environments, and they allow an experimentalist to accurately predict error suppression performance and select an optimized sequence given ambient conditions.   Specifically, we have identified the filter roll-off and ripple as limiting performance for various DD sequences, and studied how changes in pulse sequence parameters such as the parity of $n$ impact the roll-off.  Our discussion also included the consideration of realistic experimental constraints such as nonzero pulse durations and novel conclusions about timing constraints, describing potential impacts on filter performance.  We note that similar general insights -- without the specific use of the filter-design framework -- have recently appeared in the literature~\cite{Alvarez2011}

We expect continued exciting developments in dynamical decoupling sequence construction as the community begins to leverage the decades of insight provided by the field of signal processing for robust filter design.  In principle, insights coming directly from the classical control community may be leveraged for the creation of new sequences in the future, but producing useful sequences in this way remains the subject of future work.  

In the perspective presented in this paper we ignored concatenated dynamical decoupling as discussed in ~\cite{Khodjasteh2005, Khodjasteh2007, Witzel2007, West2009, Lyon2010}.  One might be tempted to assume that using local optimization techniques one could always find pulse sequences that can outperform any CDD scheme for a given number of pulses.  However, it is important to remember that with increased pulse numbers the optimization burden also rapidly increases. Therefore, new efforts focused on the creation of optimized pulse sequences for the robust preservation of arbitrary qubit states will likely incorporate optimized sequence construction for each level of a concatenated sequence as suggested in Refs.~\cite{UhrigConcatenated, FongPRL2010}.  The combination of these two approaches promises to provide vital capabilities for the implementation of error-resistent quantum memory.

Future work will also include the transfer of insights gained from dynamical decoupling sequence construction for implementation of the identity operator towards dynamical error suppression during the application of nontrivial logic -- dynamically corrected gates~\cite{ViolaDCG1, ViolaDCG2, ViolaDCG3, Young2010}.  Combined with new enhanced experimental capabilities, we believe efficient dynamical error suppression will become a foundational underlying functionality in many quantum coherent technologies.

\ack
The authors wish to thank K. Khodjasteh and W.D. Oliver for useful discussions.  MJB acknowledges support from the US Army Research Office under Contract Number  W911NF-11-1-0068, and the University of Sydney, School of Physics.  ACD supported by an Australian Research Council Future Fellowship.  This research was conducted in part by the Australian Research Council Centre of Excellence for Engineered Quantum Systems CE110001013.

\end{document}